\documentclass{article}

\usepackage{arxiv}

\usepackage[utf8]{inputenc} 
\usepackage[T1]{fontenc}    
\usepackage{hyperref}       
\usepackage{url}            
\usepackage{booktabs}       
\usepackage{amsfonts}       
\usepackage{nicefrac}       
\usepackage{microtype}      
\usepackage{lipsum}
\usepackage{graphicx}
\usepackage[ruled,vlined]{algorithm2e}

\usepackage{float} 
\usepackage{nth} 
\usepackage{chngcntr} 

\usepackage[dvipsnames]{xcolor}
\definecolor{dblue}{rgb}{0.1, 0, .80}
\usepackage{hyperref}
\hypersetup{colorlinks=true, breaklinks=true, urlcolor= dblue,
  citecolor= dblue, linkcolor=dblue}

\usepackage[english]{babel}
\usepackage[backend=biber,natbib=true,style=apa]{biblatex}
\usepackage{amsthm}
\newtheorem{theorem}{Theorem}

\graphicspath{ {./images/} }

\title{The Cluster Depth Tests: Toward Point-Wise Strong Control of the Family-Wise Error Rate in Massively Univariate Tests with Application to M/EEG}

\author{
 Jaromil Frossard \\
  Methodology and Data Analysis, Department of Psychology, \\ University of Geneva\\
  Geneva, Switzerland \\
  \texttt{jaromil.frossard@unige.ch} \\
  \AND
  Olivier Renaud\\
  Methodology and Data Analysis, Department of Psychology, \\ University of Geneva\\
  Geneva, Switzerland \\
  \texttt{olivier.renaud@unige.ch} \\
}

\newcommand{\Prob}{\mathrm{Prob}}

\begin{document}
\maketitle

\begin{abstract}
The cluster mass test has been widely used for massively univariate tests in M/EEG, fMRI and, recently, pupillometry analysis. It is a powerful method for detecting effects while controlling weakly the family-wise error rate (FWER), although its correct interpretation can only be performed at the cluster level without any point-wise conclusion. It implies that the discoveries of a cluster mass test cannot be precisely localized in time or in space. We propose a new multiple comparisons procedure, the cluster depth tests, that both controls the FWER while allowing an interpretation at the time point level. We show the conditions for a strong control of the FWER, and a simulation study shows that the cluster depth tests achieve large power and guarantee the FWER even in the presence of physiologically plausible effects. By having an interpretation at the time point/voxel level, the cluster depth tests make it possible to take full advantage of the high temporal resolution of EEG recording and give a precise timing of the start and end of the significant effects.
\end{abstract}


\section{Introduction}\label{sec:intro}

\cite{bullmore_global_1999} introduced the cluster mass test for fMRI data (functional magnetic resonance imagery) and \cite{maris_nonparametric_2007} adapted it for EEG (electroencephalography) data. It has since been adopted for pupillometry analysis \citep{hochmann_invariance_2014}. In all these cases, the statistical problem is the same: we want to test the difference between two or several conditions on signals, usually originated from measures on participants of an experiment. It can be one-dimensional signals – in time – for one EEG channel or pupillometry, or multidimensional signals – in space and time – in fMRI or multi-channel EEG. We are interested in assessing the differences between experimental conditions without any prior information on the location or on the timing of the potential true effects. In addition, the number of locations where there might be a difference is usually large (e.g., 600 time points $\times$ 128 channels in multi-channel EEG), and each corresponds to one hypothesis. From the nature of the data, the true effects are usually adjacent, in space and/or in time. This setting is often called ``massively univariate tests'' \citep{luo_diagnosis_2003} as it involves many tests using the same design. Statistically, it is part of the multiple comparisons problem.

\subsection{Classification of Tests and Error Rates}

In the multiple comparisons problem, it is useful to classify the result of a procedure on $m$ hypotheses in a $2\times 2$ table (Table~\ref{tab:classific}) similarly to \cite{benjamini_controlling_1995}. Let $R$ be the number of hypotheses called significant by the procedure, comprising $S$ correctly rejected hypotheses (also called true discoveries or true positives) and $V$ falsely rejected hypotheses (also called false positives, false discoveries or type I errors). Among the non-significant tests, $U$ are correctly non-rejected (or true negatives) hypotheses and $T$ are false negatives (or type II errors) hypotheses. Based on Table~\ref{tab:classific}, the first characterization of a multiple comparisons procedure is its family-wise error rate (FWER), which is the probability of making at least one false positive, i.e., $\Prob \left(V >0\right)$.

\begin{table}[H]
\centering
\caption{Classification of the result of a multiple testing procedure, as in \cite{benjamini_controlling_1995}. We are testing $m$ hypotheses, of which $m_0$ are under the null. We declare $R$ hypotheses significant, including $S$ correctly rejected and $V$ false positives. Among the non-significant tests, $U$ are correctly non-rejected and $T$ are false negatives.}\label{tab:classific}

\begin{tabular}{cccc}
\hline 
& non-significant & significant & \\ 
\hline 
True $H_0$ & $U$ & $V$ & $m_0$ \\ 
True $H_A$ & $T$ & $S$ & $m-m_0$ \\ 
\hline 
 & $m-R$ & $R$ & $m$ \\ 
\hline 
\end{tabular} 
\end{table}

A multiple comparisons procedure is said to control weakly the FWER if it guarantees that $\Prob \left(V >0 \right| m = m_0)=\alpha_{\rm FWER}$, where the probability is computed under the ``full null''  hypothesis, e.g.,\ under the situation where all null hypotheses are true \citep{hochberg_multiple_1987}. Otherwise, if merely one of the $m$ hypotheses is under the alternative, the procedure does not guarantee a control of the false positive for the remaining $m-1$ hypotheses. If a procedure controls only weakly the FWER, true discoveries are often coupled with false positives.

By contrast, a multiple comparisons procedure is said to control strongly the FWER if it guarantees that $\Prob \left(V >0 \right)=\alpha_{\rm FWER}$ for all possible combinations of true null and true alternative hypotheses \citep{hochberg_multiple_1987}. The strong control of the FWER implies a fixed maximal error rate for any configurations in the underlying process. It provides therefore an ideal control, as we do not know the suitable configuration. The well-known Bonferroni correction \citep{dunn_estimation_1958}  and its step-down version, the Holm correction \citep{holm_simple_1979}, both control strongly the FWER. However, to achieve this goal with a universal simple procedure, they use such loose inequalities that for massive correlated tests the actual FWER is much lower that the target value of $\alpha_{\rm FWER}$ and the statistical power is disastrous \citep{groppe_mass_2011}. 

The max-\textit{T} and min-\textit{p} procedures from \cite{westfall_resamplingbased_1993} or the Troendle method proposed by \cite{troendle_stepwise_1995} control strongly the FWER as well. They all use re-sampling by permutation to obtain the join distribution of all $m$ test statistics. The max-\textit{T} uses the test statistic itself while the min-\textit{p} uses its associated $p$-value, which avoids loss of power if the univariate distributions are different. Troendle method can be viewed as its step-down version, which proceeds by eliminating each time-point once treated, from the most to the least significant. It has uniformly a larger power than the min-$p$ method.
Since these methods include the correlation between the tests (which is preserved by permutation), they can adapt the correction to achieve a FWER at the target value of $\alpha_{\rm FWER}$. In massively univariate tests, these procedures are therefore far more powerful than Bonferroni and Holm. Quite importantly, they allow us to interpret each time point significance and therefore provide information on the precise timing of the start of a significant difference.

Although these procedures factor in the correlation between tests, they do not take advantage of the specific topology or adjacency of massively univariate tests. In that setting, the cluster mass test \citep{bullmore_global_1999,maris_nonparametric_2007} aggregates the univariate tests on the basis of spatial or temporal adjacency, which proves to drastically increase the statistical power. It should be noted that this procedure controls only weakly the FWER. It is implemented in MNE \citep{gramfort_mne_2014}, EEGLAB \citep{delorme_eeglab_2004} or permuco \citep{frossard_permuco_2018} and is also based on permutation tests.

The cluster mass test procedure is fully described in Section~\ref{sec:clustermass}. Its output declares zero, one or several significant cluster(s). A cluster is a set of adjacent time points that are aggregated to increase the power of the test. However, the interpretation of a significant cluster is not straightforward, and we cannot interpret each time point separately (point-wise interpretation). Formally, the cluster mass test only allows us to declare that we take an $\alpha_{\rm FWER}$ risk when claiming ``there is at least one alternative hypothesis within the significant cluster''. Subsequently, if we would allow ourselves to reject the null hypothesis for all time points within a significant cluster, we take a larger (and unknown) risk of a false positive than the preselected FWER target $\alpha_{\rm FWER}$. In other words, a significant cluster that contains true positives has also a large probability to include false positives as well. This is the reason, despite its many advantages, the cluster mass test cannot be used to time or to precisely localize effects \citep{sassenhagen_clusterbased_2019,groppe_mass_2011}. In particular, it would be incorrect to interpret or state that the first time point of a significant cluster is the first time the conditions differ.

The TFCE (threshold-free cluster enhancement) \citep{smith_thresholdfree_2009} is another concurring multiple comparisons procedure based on clusters that also controls weakly the FWER. It is a transformation of the statistical signals that uses the concept of cluster extent and cluster height and has the advantage of not relying on a preselected threshold parameter. Although the TFCE provides a $p$-value for each time point, it would be incorrect to interpret these $p$-values independently, and again, it would be incorrect to interpret or state that the first time point with a significant $p$-value is the first time the conditions differ.

Another approach that is very promising in fMRI concerns methods that control errors over any set of hypotheses, especially the All-Resolution Inference \citep{rosenblatt_all-resolutions_2018} and the Post-Hoc Confidence Bounds on False Positives Using Reference Families \citep{blanchard_post_2020}. These methods allow inference on any set of hypotheses that can be joint (cluster) or disjoint, while controlling for any post-hoc, user-defined selection of the set. All-Resolution Inference even provides a lower bound on the number of significant tests within the set. As with the cluster mass test, however, these methods do not provide individual tests for each hypothesis but inference at the set/cluster level. For M/EEG, they can therefore not provide information on the precise timing of the start of a significant difference.


In this article, we propose a new method of multiple comparisons that combines both the idea of clusters and the algorithm of \cite{troendle_stepwise_1995} in an effort to borrow the strength of both approaches. The aim is to provide a powerful procedure that allows us to interpret the significance of each time point (point-wise interpretation). Using simulation, we show that our new procedure corrects the excess of false positives that would occur in a point-wise interpretation of the cluster mass test while being more powerful than the min-\textit{p} or the approach from \cite{troendle_stepwise_1995}. This procedure, which we called the ``cluster depth tests'', allows researchers to interpret the results in a point-wise fashion in a much more reliable way. In particular, it implies that it can be used to precisely time the appearance of an effect.

In the next section, we present the statistical models for the massively univariate tests and the cluster mass test, before explaining why the point-wise interpretation of significant clusters creates an excess of false positives when true effects are present in the signal. Then, we present our new algorithm and show how it is related to both the cluster mass test and the algorithm from \cite{troendle_stepwise_1995}. We show the conditions for a strong control of the FWER and present a generalization to multi-channel data. Finally, we present some simulations and an application showing the advantages and limitations of our new method. The proposed method is available in the free R software \citep{rcore_r_2020} using the permuco package \citep{frossard_permuco_2018} for single-channel and a  port in python MNE package \citep{gramfort_mne_2014} is under way. The multi-channels extension is available in the \textit{permuco4brain} R package (see Section~\ref{sec:sim}).

\section{Massively Univariate Tests in General Linear Model}

In the simplest cases, we want to test the difference between experimental conditions on $m$ dependent variables. These $m$ dependent variables are distributed in time and/or in space. The present article focusses only on one-dimensional signals (e.g.,~one EEG channel), so the $m$ dependent variables are the responses for the $m$ time points. We write the models as follows:

\begin{equation}\label{eq:model}
Y_s = X\beta_s +\epsilon_s ~\forall ~s \in \{1, \dots, m\}
\end{equation}

where $Y_s$ is the response variable vector for time point $s$, $X$ is the design matrix shared by the $m$ time points, $\beta_s$ is the parameter vector of interests of length $q$ and, finally, $\epsilon_s$ the error terms. We assume that each $\epsilon_s$ follows an unknown, exchangeable distribution. Importantly, $\epsilon_s$ may be correlated across the $m$ tests. Equation~\ref{eq:model} encompasses models for analyses like regression, $t$-test or one-way ANOVA. In all these models, the tests of interest can usually be viewed as testing simultaneously $q-1$ contrasts:

\begin{equation}\label{eq:hypothesis}
H_{0:s}:~ G\beta_s = 0,
\end{equation}

where $G$ is a $q-1\times q$ contrast matrix. The $F$ or $t$ statistic are typical choices as the univariate test statistic for each $m$ time point, but the procedures can accept any statistic. Let $F_s$ be the statistic at time point $s$. The theory behind all multiple testing procedures is based on the (unknown) joint distribution of the statistics $[F_1 ~\dots ~F_m]^\top$. If the experiment contains more that one factor, the methods presented in \cite{winkler_permutation_2014} \cite{kherad-pajouh_exact_2010} and \cite{frossard_permuco_2018} to test any main or interaction effect can easily be extended to cover all multiple comparisons procedures presented here, including the cluster depth tests. The latter reference includes methods for cases with between- and/or within-subject factors.

\section{False Discoveries in the Cluster Mass Test}\label{sec:clustermass}

The cluster mass test is well described by an algorithm. First, one computes univariate statistics (e.g., $t$ or $F$ statistic) for each time point of the signals to produce a statistical signal (or statistical map in higher dimensions). Then, one uses a predefined threshold ($\tau$ usually set at the \nth{95} percentile of the parametric null distribution for the univariate statistic) to construct clusters on the statistical signals as follows. All adjacent time points whose statistics are above the threshold create together one cluster. For each cluster, the cluster mass is computed by aggregating the univariate statistics using their sum (or sum of squares)\footnote{The length of the cluster can also be used (and can be viewed as a function of the statistics).}. The $p$-values are computed for each cluster by comparing its cluster mass with the cluster mass null distribution. This distribution is computed using permutation: we permuted the signals and repeated the algorithm, i.e., we compute univariate statistics on the permuted signals, create clusters using the threshold, and compute the cluster mass for each cluster. For each permuted dataset, we keep the maximal cluster mass (if multiple clusters are found). The maximal cluster mass for all permuted datasets form together the null distribution of the cluster mass \citep{maris_nonparametric_2007}.

The cluster mass test should only be interpreted at the cluster level. One correct interpretation is ``We take a risk of probability $\alpha_{\rm FWER}$ when claiming, there is at least one true effect within a significant cluster''. This implies that we cannot use the cluster mass test to describe with high precision a cluster’s size and location (shape in higher dimensions). Formally, the concept of a false positive is not well defined in the cluster mass test as we do not have the statistical decision for each time point, although the null hypotheses of Equation~\ref{eq:hypothesis} are defined at each time point.

\begin{figure}[H]
\includegraphics[width=\textwidth, height=\textheight, keepaspectratio]{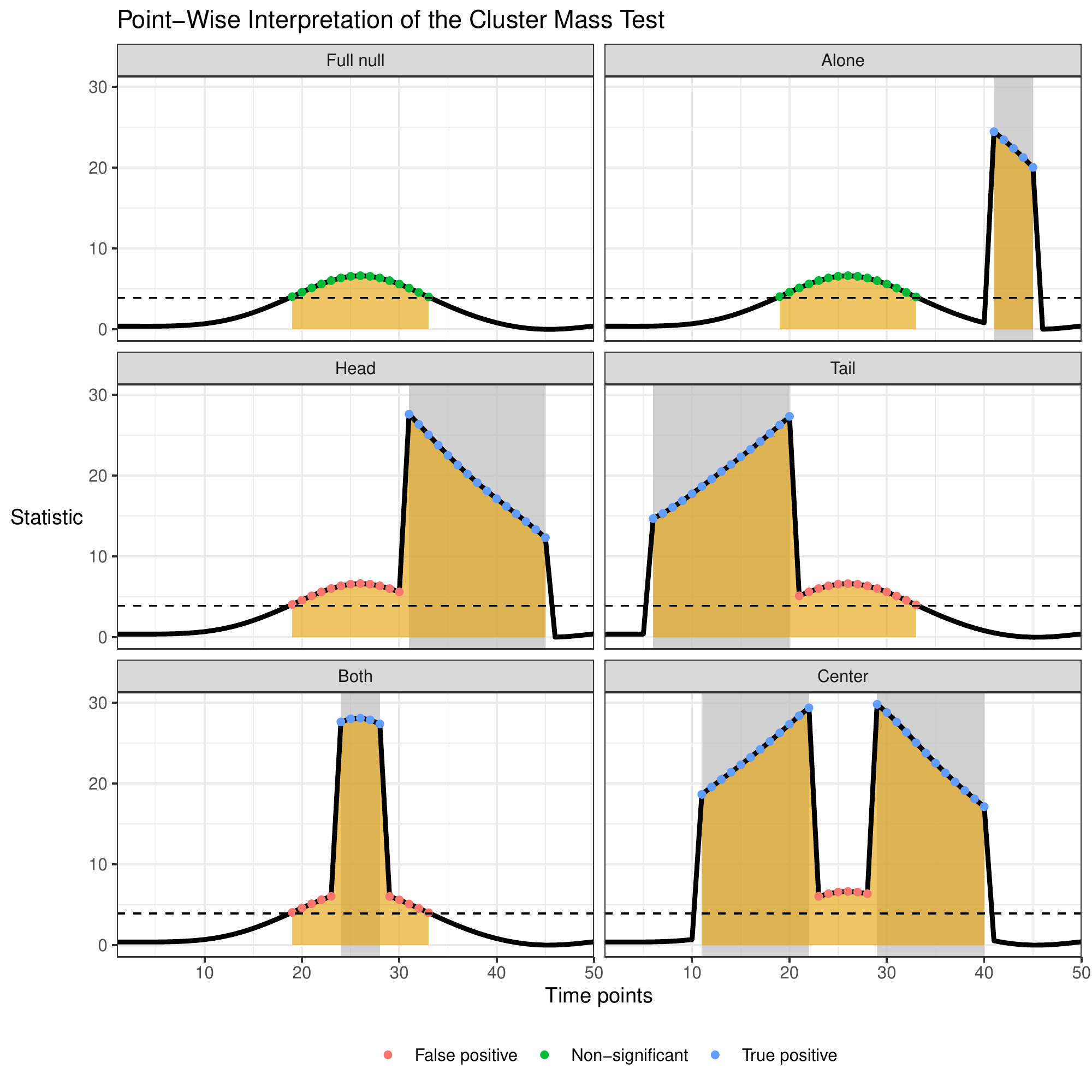}
\caption{Six cases where regions of true effects (gray area) combine with a cluster created by noise only. The gray area indicates the presence of a region of true effect, and the orange area indicates clusters created using the predefined threshold (dashed line). In the four panels ``Head'' , ``Tail'' , ``Both''  and ``Center'', the point-wise interpretation of the cluster mass test leads to false positives as some null hypotheses will be part of clusters driven by a true effect.}\label{fig:cm_problem}
\end{figure}

In order to understand how the cluster mass test may be improved to allow a point-wise testing, up until the next section we will purposefully over-interpret the results of a cluster mass test with a point-wise interpretation (i.e.,\ all time points within a significant cluster are interpreted as individually significant). In Figure~\ref{fig:cm_problem}, we show six cases where we observed the same noise but different true effects. In the ``Full null''  panel, we observed one cluster (of noise) with a small cluster mass. The cluster mass test takes only a risk $\alpha_{\rm FWER}$ to call a significant cluster under the full null hypothesis, so this cluster is very likely to be deemed non-significant. In the ``Alone''  panel, we observed a second cluster that is produced by a true effect (gray region) that is significant. In the first two cases, no false positive is perpetrated. In the ``Head''  panel, a single significant cluster results from the merger of two other clusters: one cluster that is formed by the region of true effect, and one cluster that appears due to noise only. Using the point-wise interpretation of the cluster mass test, all time points within this single significant cluster have the same cluster mass, which implies the same statistical decision. It produces therefore several false positives at the head  of the cluster. In the ``Tail''  panel, the tail  of the observed cluster is affected by false positives, while in the ``Both''  panel both the head  and the tail  of the cluster are affected by false positives. Finally, in the ``Center''  panels, we discover one significant cluster that is the aggregation of two regions of true effect linked by noise. In practice, true cases similar to the ``Head'', ``Tail''  or ``Both''  panels are not uncommon. It depends on the correlation structure of the noise, but more importantly on the chosen threshold $\tau$ and on the number of ``borders''  of a region of true effect. Hence, the value of the statistic for the time point (under the null hypothesis) at the border of a region of true effect has a probability near $5\%$ to be above the threshold (when the threshold is set at the $95\%$ of the null distribution of the parametric statistic). In this case, the observed cluster contains both time points under the null and alternative hypotheses. The above probability increases as a true effect has two borders for a one-dimensional signal and even more in higher dimensions. When the power of the test is large, more than $5\%$ of significant clusters produced by regions of true effect contains at least one false positive \footnote{This is an excess of false positives, as, even when some hypotheses are under the alternative, there might be significant clusters that are fully composed of null hypotheses. It would be less than $5\%$ as we assume a true effect ($m_0 < m$).}. This is a lower bound that suffers greatly from the ``curse of dimensionality''  as this probability depends on the number of time points at the border of a region of true effect.

This excess of false positives comes from the aggregation of statistics within the clusters. All time points of a cluster are associated to the same cluster mass: at the border of a region of true effect, some time points under the null hypothesis might ``benefit''  from the contribution of a true effect as they will all share the same cluster mass. 
The aim of the cluster depth tests is to take into account this excess of false positives at the head  and tail  of clusters. They re-use the concept of clusters created by a threshold while dropping the concept of cluster mass to reach point-wise tests within the clusters.

\section{Point-Wise Significance within the Clusters: the Cluster Depth Tests}\label{sec:samplewise}


The cluster mass test is general to cases where hypotheses are distributed in time and/or in space. It only needs that some suitable adjacency between points can be defined. In this article, we restricted ourselves to one-dimensional signals as we rely on the concept of cluster depth. The cluster depth describes the position of a time point relative to the cluster’s first or last time point. Each time point within a cluster is therefore associated with two cluster depths, one from the cluster’s head and one from its tail.

\subsection{The Cluster Depth Tests Algorithm}
 
If the time point $s^\ast$ is the first time point of an observed cluster, or at the first cluster depth from the head, the ``classical approach'' to testing its corresponding hypothesis would be to compare the observed statistic at time point $s^\ast$ to the null distribution at that specific time point $s^\ast$, while taking into account that $m$ tests are carried out (one for each time point). In that procedure, the time point $s^\ast$ is relative to the start of the signal. Instead, we propose to compare the statistic at time point $s^\ast$ relatively to its position within the cluster. In that case, it would be the first cluster depth (as we assumed that $s^\ast$ corresponds to the first time point of an observed cluster). We produce a test for the first cluster depth as follows. First, we compute the statistic of interest for each time point of the signals; then, using the predefined threshold $\tau$, we create clusters. If needed, we remove the cluster that includes the first time point as we cannot deduce the cluster depth (from the head) of its individual statistics. The first cluster depth statistic (from the head) is the statistic of the first time point within the cluster. We produce a $p$-value for the first cluster depth by comparing its observed statistic with its distribution under the null hypothesis. This distribution is computed by permutation: for each of the $n_P$ permuted datasets, we compute the univariate statistic at each time point; then, using the threshold $\tau$, we create clusters, again discarding a cluster that would start on the first time point. If several clusters are created, we keep the maximal values at the first cluster depth;  if there is no cluster for a given permutation, this value is set to 0. These $n_P$ maximal values create together the null distribution by permutation of the first cluster depth. Algorithm~\ref{algo:first} describes this procedure, which provides a $p$-value for each time point associated to the first cluster depth. The distribution by permutation of the first cluster depth is interpreted as the distribution of the first value of the clusters given (1) no difference between conditions at the first cluster depth, (2) signals of length $m$ and with stationary errors, and (3) clusters created using a threshold $\tau$. Observing a large value of the statistic of first cluster depth indicates an uncommon observation under the null hypothesis for this first value. So, we would take an $\alpha$ risk rejecting the null hypothesis for the test associated to the first cluster depth when its $p$-value is inferior to $\alpha$.\\

\begin{algorithm}[H]
\SetAlgoLined
\KwResult{$K$ $p$-values associated to the $K$ time points for the first cluster depth (of the $K$ observed clusters).}

Compute the univariate statistics for the $m$ tests for the original data and $n_P$ permutations. It corresponds to $n_P$ statistical signals $\tilde\mathcal{F}_{0:[i]}$ of length $m$ that are stored in a $n_P\times m$ matrix $\tilde\mathcal{F}_0$.\\

\For{each $i \in \{1,\dots,n_P\}$}{
 
Using the threshold $\tau$, compute $K^*$ clusters on the permuted signal $\tilde\mathcal{F}_{0:[i]}$ as sets of adjacent time points whose statistics are above the threshold and do not include the first time point. \\

If $K^*=1$, return the statistic of the first cluster depth.\\

If $K^*>1$, return the maximal statistic of the first cluster depth over the $K^*$ clusters.\\

If $K^*=0$, return $0$.

}

The $n_P$ returned values constitute the null distribution of the first cluster depth ($D_{\textrm{depth: }1}$) statistic.

Compare ($D_{\textrm{depth: }1}$) to the observed statistics at the first cluster depth of the $K$ observed clusters to produce $K$ $p$-values.
\caption{Inference for the first cluster depth from the head.}\label{algo:first}
\end{algorithm}

Note that Algorithm~\ref{algo:first} can be viewed as a special case of the cluster mass test. Indeed, the cluster mass test uses a function to aggregate the individual statistics into a cluster mass statistic. It is usually the sum or the sum of squares of the statistics of all time points within the cluster. However, defining this function as a weighted sum, with a weight of 1 for the first time point of the cluster and a weight of 0 otherwise, gives the same procedure as Algorithm~\ref{algo:first}. Moreover, contrary to the cluster mass statistic, the cluster depth statistic of a null effect will not be influenced by a true effect that appears after the cluster depth of interest.

\begin{figure}[H]
\includegraphics[width=\textwidth, height=\textheight, keepaspectratio]{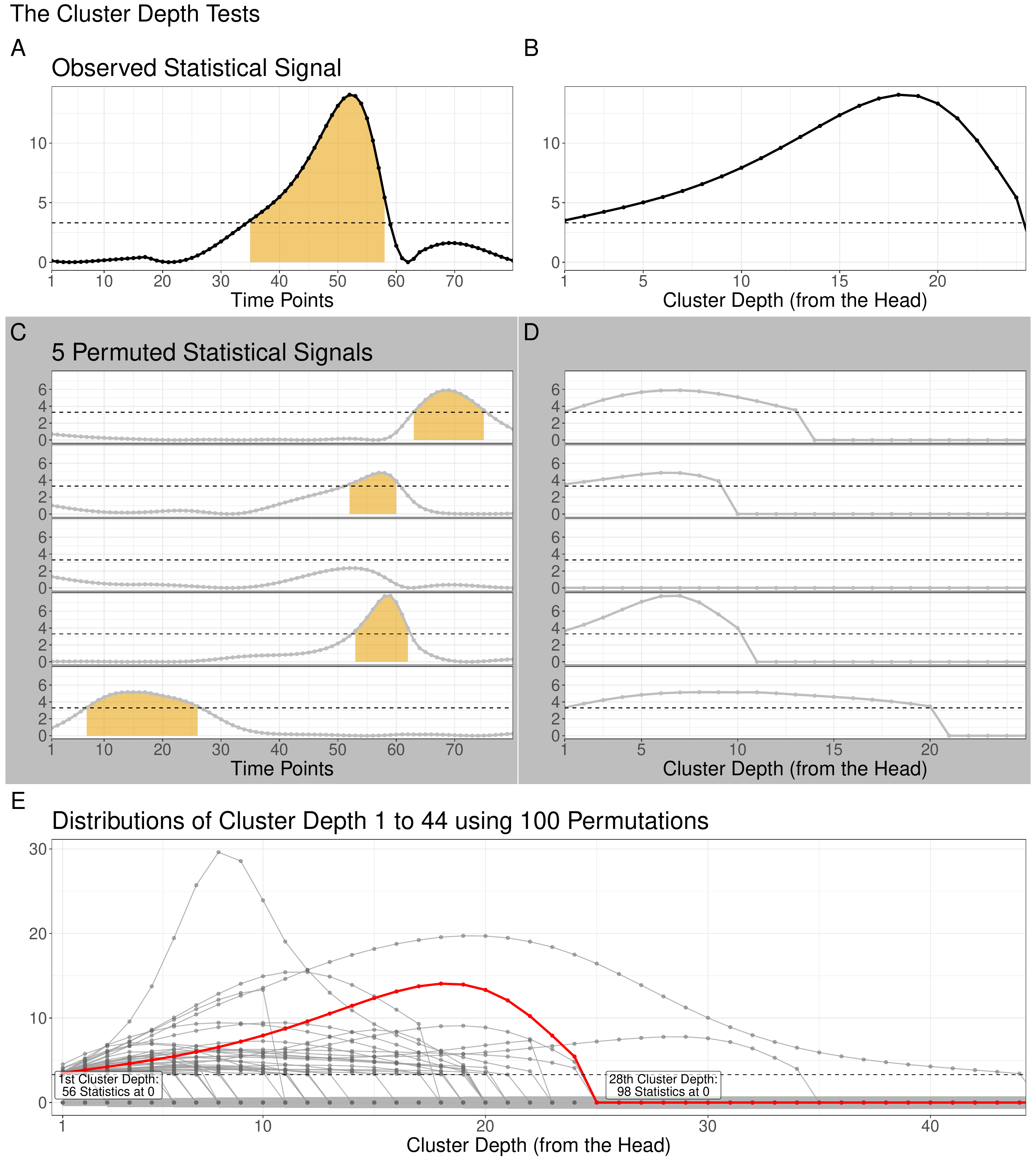}
\caption{Distribution of the cluster depth statistics (from the head ) as described in Algorithm~\ref{algo:all}. Panel~A shows the observed statistical signal, which contains one cluster. In Panel~B, we extract the cluster of length 24. Panel~C shows five permuted signals. For each permuted signal, we computed clusters (orange) and re-align them according to their cluster depth (Panel~D). In Panel~E, for each cluster depth the observed statistic (in red) and its distribution under the null hypothesis (formed by all dots in gray) are shown. For depth larger than 44, the distribution is degenerated with a point mass in 0. Finally, using the method from \cite{troendle_stepwise_1995}, we test individually each cluster depth while taking into account all cluster depths.}\label{fig:depth_head}
\end{figure}

The next step is to generalize Algorithm~\ref{algo:first} to all cluster depths and is described in Algorithm~\ref{algo:all}. There are $m-1$ cluster depths as we discard clusters that begin on the first time point. Given that we observed $k\in \{1,\dots,K\}$ clusters of length $J_k\in\{J_1,\dots,J_K \}$, we compare the observed statistic at the $j$\textsuperscript{th} cluster depth to its distribution under the null hypothesis for each $j$\textsuperscript{th} cluster depth $j \in \{1,~\dots,~ m-1\}$. The null distribution of the $j$\textsuperscript{th} cluster depth is computed by permutation where, for each permutation, we use the threshold to form clusters discarding eventual clusters that start at the first time point. Then, for each permutation, we keep the statistic at the $j$\textsuperscript{th} cluster depth. If there are multiple clusters whose length is at least $j$, we keep the maximal statistic of the $j$\textsuperscript{th} cluster depth, and if no cluster is found or the maximal length of the clusters is below $j$, then the value is set to $0$. This procedure creates by permutation the $m-1$ cluster depth distributions. Note that for the largest cluster depths such that $j>J_D$, where $J_{D}$ corresponds to the length of the longest cluster among the observed or permuted statistics, the distributions are degenerate with a point mass at 0. Advanced multiple comparisons procedures like Troendle will allow us to discard them. The distributions of the cluster depths presented in Algorithm~\ref{algo:all} are illustrated in Figure~\ref{fig:depth_head} using a simple example. These distributions are highly correlated due to the adjacency of the depths, and the permutation procedure captures correctly their full multivariate distribution. It would inflate the FWER to directly compute and interpret the resulting $J$ univariate $p$-values, since potentially $m-1$ tests are carried out. As Troendle's procedure computes the $p$-values sequentially, always looking at the smallest values, only the non-degenerate distributions influence the procedure, and we can discard all but the first $J_D$ distributions. The \cite{troendle_stepwise_1995} method has proven to be powerful when comparing a reasonable number of tests, and the proposed method benefits from it in controlling the FWER among the $J_D$ cluster depth tests while maintaining a high power. Note that the lengths $J_k$ of the observed clusters are at most $J_D$, and we are only interested in the first $J_k$ $p$-values.

\begin{algorithm}[H]
\SetAlgoLined
\KwResult{$p$-values associated to all time points within the $K$ observed clusters.}

Compute the univariate statistics for the $m$ tests for the original data and $n_P$ permutations. It corresponds to $n_P$ statistical signals $\tilde\mathcal{F}_{0:[i]}$ of length $m$ that are stored in a $n_P\times m$ matrix $\tilde\mathcal{F}_0$.\\

Using the threshold $\tau$, define $K$ clusters on the observed statistical signal as the sets of adjacent time points whose statistics are above the threshold and do not include the first time point. The length of the $k$\textsuperscript{th} cluster  is $J_k$, for $k=1,\dots,K$.

\For{each $j$\textsuperscript{th} cluster depth in $1,\dots,m-1$}{
Define $D_{\textrm{depth: }j}$ as the null distribution of the statistic at the $j$\textsuperscript{th} cluster depth as follows:\\

\For{each $i \in \{1,\dots,n_P\}$}{

Using the threshold $\tau$, compute $K^*$ clusters on the signal $\tilde\mathcal{F}_{0:[i]}$ as sets of adjacent time points whose statistics are above the threshold and do not include the first time point. \\

If $K^*=0$ or if the maximal length of all $K^*$ clusters is less that $j$, return 0 for this permutation.

If $K^* =1$, return the statistic at the $j$\textsuperscript{th} cluster depth\\

If $K^* >1$, return the maximal statistic at the $j$\textsuperscript{th} cluster depth over the clusters whose length is greater than $j$.
}
The $n_P$ returned values constitute the null distribution of the statistic at the $j$\textsuperscript{th} cluster depth : $D_{\textrm{depth:}j}$.

}

Keeping the same order of the permutations, the $m-1$ vectors of length $n_P$ constitute the multivariate null distribution of the cluster depths. Let $J_D$ be the last cluster depth which distribution is not degenerate (i.e. the largest cluster length among all permutations). We store the $J_D$ distributions in a $n_P\times J_D$ matrix $[D_{\textrm{depth:}1},\dots, D_{\textrm{depth:}J_D}]$.

Compare this multivariate distribution to the observed statistics of the $K$ clusters in order to produce $\sum_k J_k$ $p$-values, which correspond to one $p$-value for each time point within the $K$ clusters, as follows:

\For{each $k=\{1,\dots,K\}$ clusters}{
 When $J_{k}< J_D$, fill the statistics of the $k$\textsuperscript{th} cluster  with 0 to obtain $J_D$ values.

Compare the $J_D$ values to the multivariate null distribution $[D_{\textrm{depth:}1}~\dots~ D_{\textrm{depth:}J_D}]$ using the algorithm proposed by \cite{troendle_stepwise_1995}. 

Keep the $J_{k}$ corrected $p$-values corresponding to each time point within the $k$th cluster.

}

\caption{Inference for all cluster depths $1,\dots,m-1$ from the head }\label{algo:all}
\end{algorithm}

Using Algorithm~\ref{algo:all} and computing the cluster depth from the head of the cluster only takes into account the false positives that may occur at this head (Figure~\ref{fig:cm_problem}, panel ``Head'' ). It already improves the interpretation of significant time points. If we observed a significant effect from time points $50$ to $70$ using the cluster depth tests (from the head ), we take an $\alpha_{\rm FWER}$ risk when claiming that ``there is a true difference that begins no later than time point 50'' . However, we take a risk larger than $\alpha_{\rm FWER}$ if we claim that the effect lasts at least to time point $70$. Due to the low power at the extremities of effects, note that the difference between experimental conditions may begin sooner than $50$ (or after $70$), but it would imply merely a false negative or negatives. 

Using Algorithm~\ref{algo:all} while counting the depth from the cluster head does not provide any correction for the false positives that may appear at the cluster tail (Figure~\ref{fig:cm_problem}, panel ``Tail'' ). However, by symmetry, we can apply Algorithm~\ref{algo:all} while counting the depth from the cluster tail (or by reversing the time). Using the cluster depth from both the head and the tail produces two $p$-values for each time point within a cluster. To protect the FWER, we keep the largest of the two to evaluate the significance of each time point, as described in Algorithm~\ref{algo:both}. The largest will typically be the head one for the cluster’s starting point and the tail one for the ending point, but the algorithm protects automatically any case.

\begin{algorithm}[H]
\SetAlgoLined
\KwResult{$p$-values associated to all cluster depths within the clusters.}
 
 Use Algorithm~\ref{algo:all} to compute the cluster depth tests from the head. It produces the $p$-values $p_H$.
 
 Use Algorithm~\ref{algo:all} while reversing the order of the time to compute the cluster depth tests from the tail. It produces the $p$-values $p_T$. Reorder $p_T$.
 
Take the element-wise maximum between $p_H$ and $p_T$, producing the $p$-values $p_B$.

Return $p_B$.

 \caption{Controlling the false positives at the head  and tail  of the clusters.}\label{algo:both}
\end{algorithm}




\subsection{Details on the Permutation Scheme}

Implementations of the cluster mass or of the threshold-free cluster enhancement (TFCE) procedures to compare experimental conditions as in Equations~\ref{eq:model} and~\ref{eq:hypothesis} usually directly permute the signals. It corresponds to the classical permutation method, usually attributed to \cite{manly_randomization_1991} for simple and multi-factorial designs. It guarantees that the multivariate distribution of the statistics is stationary only under the assumption of the full null hypothesis (all $m$ hypotheses in Equation~\ref{eq:hypothesis} are true) and of the stationarity of the error terms. By contrast, for a strong control of the FWER, a procedure must work when any set of the time points is under the alternative. In this situation, directly permuting the signals do not provide statonarity, not even asymptotically, for the following reason: if the statistics are based on permuted raw data, the correlation between two statistics (for two different time points) will be highly influenced by the true effect at these two time points (i.e. by the $\beta_s$), implying that the statistical signal is not stationnary.

Fortunately, a simple solution exists. It consists in removing the observed averages per group before permuting the signals such that the re-samples reflect the full null hypothesis. For the simple model of Equations~\ref{eq:model} and~\ref{eq:hypothesis} it is sufficient to make the permuted signals asymptotically stationary (see the proof of Theorem~\ref{thm:fwer}). It is the same idea as in the ``bootstrap-t'' used in LIMO EEG \citep{pernet_limo_2011}. For more general multi-factorial designs and hypotheses, this procedure has already been generalized both for permutation and for the bootstrap by \cite{terbraak_permutation_1992}. In that case, the re-samples are constructed as the fitted values plus a re-sampling of the full residuals, i.e., the residuals of the full model and the hypothesis tested is shifted to the observed value \citep{winkler_permutation_2014,frossard_permuco_2018}. As in these references, we will call it the ter Braak permutation scheme. In the simulations of Section~\ref{sec:sim}, we evaluated the cluster depth tests using either the ter Braak or the Manly permutation method, and both showed very similar results concerning the FWER.

Finally, when the number of observations becomes moderately large, it is not feasible to compute the permutation test on a modern computer using all permutations ($n_P$ in Algorithm~\ref{algo:first}~and~\ref{algo:all}). However, we can estimate the permuted $p$-values using a large enough number of permutations. Given the confidence intervals on the permuted $p$-values presented by \cite{winkler_faster_2016}, we recommend to use at least a few thousand randomly selected permutations.


\subsection{Strong Control of the FWER and Limitations}

\begin{theorem}\label{thm:fwer}
 Under the model in Equation~\ref{eq:model} and under stationarity of the errors $\epsilon_s$, to test the hypotheses in Equation~\ref{eq:hypothesis} the procedure described in Algorithms~\ref{algo:all} and \ref{algo:both} where
  \begin{itemize}
  \item the group averages are removed before the permutation (or more generally, the \cite{terbraak_permutation_1992} permutation method is used) and
    \item a $t$ or a $F$ statistic is used (an asymptotically pivotal statistic)   
  \end{itemize}
  asymptotically controls strongly the FWER for all clusters that contain at most one bloc of true effects.
\end{theorem}

The proof can be found in Appendix~\ref{sec:theoryappendix}. Its idea is that for time points prior to the bloc of true effects, only the leading time points will matter, while for time points after the bloc of true effects, only the trailing time points will matter. Since all these points are under the null, the construction of the join distribution of cluster depths under the null (from the head or respectively from the tail) and its correction for multiple tests will control the type I errors.

Note that the need of stationarity is also true for the cluster mass test and the TFCE, but they both give guaranties only under the full null hypothesis, i.e., when all $m$ hypotheses in Equation~\ref{eq:hypothesis} are true.

For models with more than one factor, the theorem holds only for the permutation method from \cite{terbraak_permutation_1992} as none of the other permutation methods presented in \cite{winkler_permutation_2014} or \cite{frossard_permuco_2018} use the full residuals. The reason is that some clusters formed on the permuted data may potentially cover more than one bloc of true effects, which implies that some values in the construction of the cluster depth statistics might be erroneous. However, in practice the results will be quite similar, except in very artificial cases where the alternative hypotheses (true effects) form a comb-like shape. Using the \cite{manly_randomization_1991} method instead of the \cite{terbraak_permutation_1992} method in the simulation below leads to extremely similar results (see the simulations in Section~\ref{sec:sim}).

To explicitely formulate its limitation, the cluster depth tests procedure does not fully control strongly the FWER as it does not guarantee, for all combinations of null and alternative hypotheses, a FWER at the nominal level. Schematically, it controls strongly the FWER in the first five cases depicted in Figure~\ref{fig:cm_problem} but might fail in the sixth case. The FWER is guaranteed therefore when the observed clusters contain one region of true effects. It is likely to happen when regions of true effects are far enough apart, which often corresponds to the expected physiologically plausible effects in M/EEG. Typically, the cluster depth tests procedure will not guarantee the FWER when two true effects are separated with only one time point under the null (see the simulations in Section~\ref{sec:sim}). This implies that the interpretation of the cluster depth tests should not be a fully point-wise interpretation of the tests. For instance, finding significant time points from $50$ to $70$ with the cluster depth tests would formally allow us to declare the following at an $\alpha_{\rm FWER}$ risk: ``There is at least one region of true effect from time points $50$ to $70$, beginning no later than time point $50$ and ending no earlier than $70$.'' However, we do not know exactly how many true region(s) there might be within the time frame. Note that the interpretation would be the same if all observed statistics from time points $50$ to $70$ are above the threshold (they form together one cluster), but only the time points from $50$ to $55$ and $65$ to $70$ are declared significant by using the cluster depth tests. In that example, only the borders at time points $50$ and $70$ are controlled for false positives. Note, however, that these mistakes that make the cluster depth tests miss the strong control of the FWER will happen quite rarely: not only must two true effects be close to one another, but in conjunction all time points in-between (that are under the null) should maintain a statistic above the threshold, which by definition of the threshold occurs in less that 5\% of the cases. Moreover, physiologically plausible effects always start and end slowly, which substantially increases the probability that at least one time point is below the threshold, cutting the cluster in two and decreasing the false positive rate.

\subsection{Extension to Multi-Channels Analysis}\label{sec:multi}

A natural extension of the cluster depth tests to multi-channels data computes the cluster depth distributions independently for each channel, and then aggregates the distributions across the channels (per cluster depth, per permutation) using the maximum. It creates a multivariate distribution that is used, in combination with the \cite{troendle_stepwise_1995} method to produce a test (see an example in Section~\ref{sec:datamultichannel}). This test controls the FWER across all channels as it uses an argument similar to the max-$T$ procedure across the cluster depth statistics. Theoretically, it is as if all channels data are bounded side by side (with a zero value in-between), which formally shows that the above theorem applies to this multi-channels extension. A more general extension to spatial data, with no time dimension (or where time is not considered as a special dimension), is more complex as the depth is more disparate in higher dimensions. With merely a time dimension, the strength of the methods comes from the fact that each time point within a cluster is associated to only two depths (from the head and from the tail). In higher dimensions, there are many paths on which a depth can be measured, which makes the extension to spatial data non-trivial.

\section{Simulation Study: FWER and Power}\label{sec:sim}

We simulate signals of length $m=400$, from two groups of 10 participants. Then, we test the differences between the two conditions with an $F$ statistic and compare several multiple comparisons procedures. We vary the correlation between the time points error terms using either an independent, a Gaussian or an exponential autocovariance function \citep{abrahamsen_review_1997}. In all simulation settings, the univariate error terms follow standard normal distribution. Moreover, we vary the number of regions of true effects (0, 1 or 2) and the number of time points under the alternative ($1\%$, $10\%$, $20\%$, $40\%$ or $60\%$ of $m$). For cases with only one region of true effects, it is centered in the middle of the signal (at time point 200). For cases with two regions, the number of time points under the alternative is first split in two regions of the same size. In the ``2 regions'' condition, the two regions are centred at one third and two thirds of the time period. By contrast, in the ``2 nearby regions'' condition, the two regions are separated by only one time point that represents the worst case for the cluster depth tests. We also vary the shape of these regions, with either a square-shaped effect region, where the $\beta_s$ are either $0$ or $\beta_{max} \neq 0$, or a region with a right-angled triangular-shaped effect where the $\beta_s$'s increase linearly from $0$ (at the beginning) to $\beta_{max} \neq 0$ at the end of the region of true effect. Finally, we modify the values of $\beta_{max}\in \{0.2,~0.8,~1,~1.5,~2\}$ to better understand its influence on the tests of regions under the null hypothesis (FWER) and on their power in the regions under the true effect. Each of the simulation settings is replicated 4000 times to evaluate the methods. We also compare the cluster depth tests with two different permutation methods. For the ``cluster depth (ter Braak)'' approach, we remove the average before permuting the data. This method generalizes to the \cite{terbraak_permutation_1992} method in multi-factorial designs and is used in Theorem~\ref{thm:fwer}. We also present the ``cluster depth (Manly)'', which is the classical permutation method where no transformation is applied before permuting the data. This second method generalizes to the \cite{manly_randomization_1991} method in multi-factorial designs.

The proposed method is compared to several approaches to control for multiple comparisons: the min-\textit{p} procedure \citep{westfall_resamplingbased_1993}, Troendle procedure \citep{troendle_stepwise_1995}, the cluster mass test \citep{maris_nonparametric_2007} with a threshold at the \nth{95} percentile of the statistic, and finally the TFCE \citep{smith_thresholdfree_2009} (with $H=1$ and $E=0.5$ as suggested in \cite{pernet_clusterbased_2015} and with the same threshold as the cluster mass test). For each multiple comparisons procedure, we compute the $p$-values using 5000 random permutations.

All simulations are performed in R \citep{rcore_r_2020} using the permuco package \citep{frossard_permuco_2018}; the cluster depth tests are currently available in the ``rcpp'' branch (\url{https://github.com/jaromilfrossard/permuco})\footnote{It can be installed using the ``remotes'' package \citep{remotes} with \texttt{remotes::install\_github("jaromilfrossard/permuco@rcpp")}}.
We measure the FWER as the proportion of simulations that exhibit at least one false discovery (with 95\% CI using \cite{agresti_approximate_1998}), i.e., for which at least one time point outside the true effect region(s) is called significant. Two aspects of power are evaluated: first, the average percentage of true discovery within the region(s) of true effect (average power), and second, the proportion of simulation where at least one time point under a region of true effects is significant (disjunctive power, \cite{bretz_multiple_2010}). The min-\textit{p} and Troendle procedures and the cluster depth tests are meant to be interpreted point-wise, whereas such a point-wise interpretation of the cluster mass and TFCE procedures would represent an over-interpretation of their results, as explained in Section~\ref{sec:clustermass}. However, for the sake of the comparison, it is of interest to see how point-wise results of all methods differ.

\begin{figure}[H]
\includegraphics[width=\textwidth, height=\textheight, keepaspectratio]{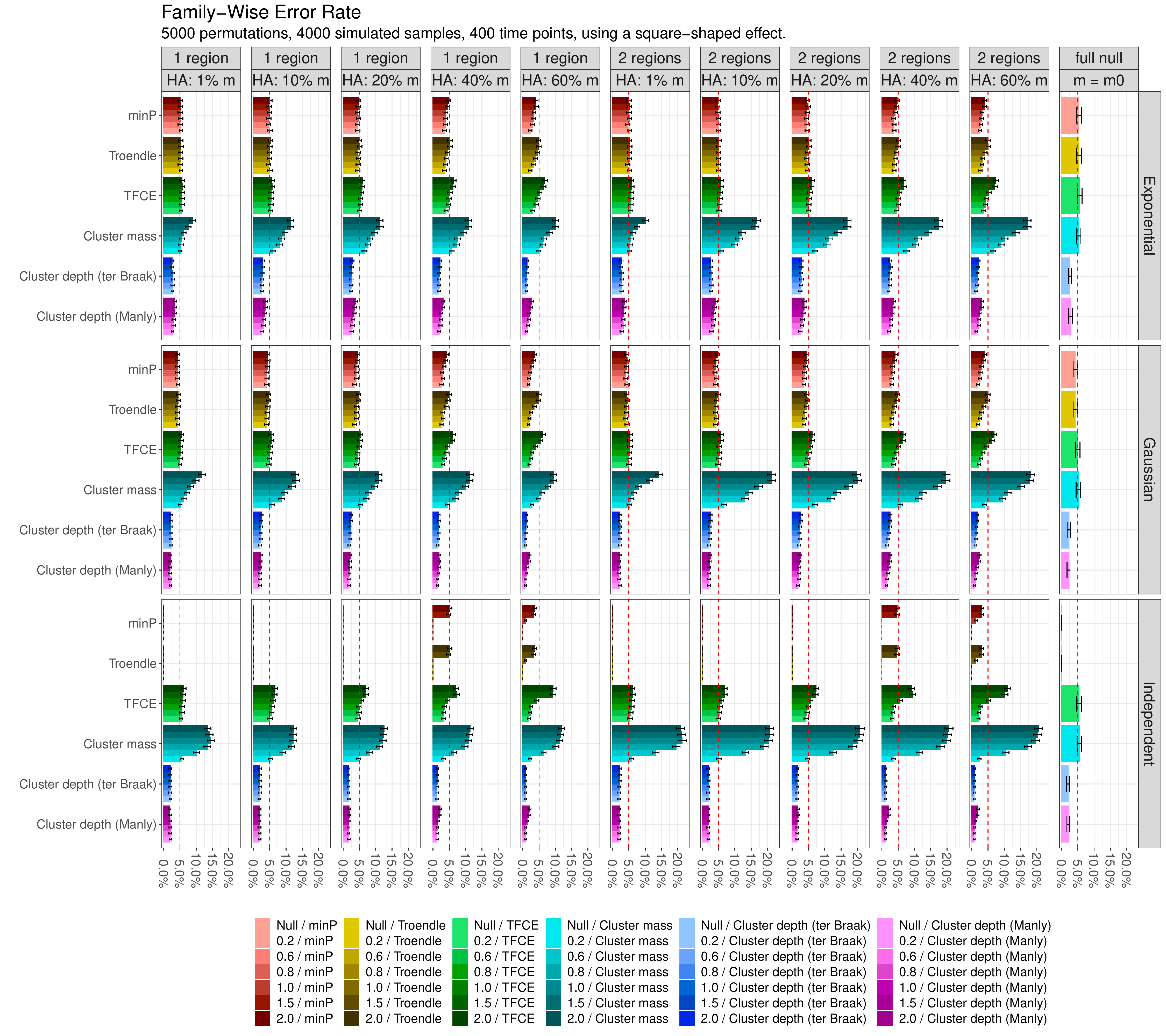}
\caption{Empirical FWER under different numbers and lengths of true effect regions (column), different noise structures (rows) for different multiple comparisons procedures (colors) and with different effect sizes of the true effects (luminance of the colors). The FWER is computed on all time points that are under the null hypothesis and should at most 5\%. Even in the presence of true effects, the cluster depth tests control the FWER, in a conservative way. The min-\textit{p} and \cite{troendle_stepwise_1995} procedures cannot handle a number of independent tests, the cluster mass test can be very anti-conservative and the TFCE can sometimes be conservative and sometimes anti-conservative.} \label{fig:fwer_sq_ok}
\end{figure}

The results of the FWER for all the settings described in the first paragraph and all the multiple testing procedures are shown in Figure~\ref{fig:fwer_sq_ok} for a square-shaped effect and in Figure~\ref{fig:fwer_tri_ok} for a triangular-shaped effect. The different columns show the results for signals with a different proportion of time points under true effect and a different number of regions of true effect. In the last column, no time points are under the true effect, i.e., the signals are under the full null ($m_0=m$). The color luminance represents the strength of the effect in the region(s) of true effect, but the FWER is of course computed in all time points that are outside each region.

A method that would strongly control the FWER would have bars that are always at the 5\% level or lower, being conservative when it is less than 5\%. Note that it would control the FWER only weakly if the bars are at the 5\% level or lower only in the last column and in the ``null''/lighter colored cases.

As expected, the min-\textit{p} and Troendle multiple comparisons procedures control strongly the FWER even in the presence of one region or two distant regions of true effect (Figure~\ref{fig:fwer_sq_ok} and Figure~\ref{fig:fwer_tri_ok}). However, we see that they are conservative with an actual FWER often even equal to 0 when the noise is independent. It can be explained by numerical issues of the min-$p$ method. Indeed, when the equivalent number of independent test is large relatively to the number of permutations, it becomes likely than more than 5\% of the permuted datasets contain at least once the largest statistic which corresponds to a ``\textit{p}-value'' equal to $1/n_P$. In this case, the min-\textit{p} distribution has a frequency larger than 5\% at $1/n_P$ and it cannot reject any hypotheses. This numerical issue can be resolved by increasing the number of permutations. Indeed, when $n_P<\tilde{n_P}$, the frequency at $1/n_P$ of the min-\textit{p} distribution (using $n_P$ ) will always be larger or equal to the frequency at $1/\tilde{n_P}$ of the min-\textit{p} distribution (using $\tilde{n_P}$). However, the minimal number of permutation needed to validly prevent this problem is \textit{a priori} unknown and increasing the number of permutations is compute-intensive especially, when the number of tests is large.


In the settings of Figures~\ref{fig:fwer_sq_ok} and~\ref{fig:fwer_tri_ok}, these simulations show that the proposed cluster depth tests control the FWER and are actually slightly conservative. Indeed, the cluster depth tests double the correction, from both the ``head'' and the ``tail'', which is, in many cases, redundant. For instance, under the full null, the correction from only one of the directions would be sufficient to guarantee the FWER. Moreover, the two versions of the cluster depth tests show similar results, although the ``Manly'' version seems slightly less conservative. On the contrary, the cluster mass test does not guarantee the FWER when true effects are present, although it controls weakly the FWER under the full null. Finally, in these simulation settings, although with a FWER sometimes significantly larger than 5\%, the TFCE is generally close to the FWER target, which is not theoretically guaranteed. This confirms the good property of an extend parameter $H<1$ as suggested by \cite{smith_thresholdfree_2009} and \cite{pernet_clusterbased_2015}.

\begin{figure}[H]
\includegraphics[width=\textwidth, height=\textheight, keepaspectratio]{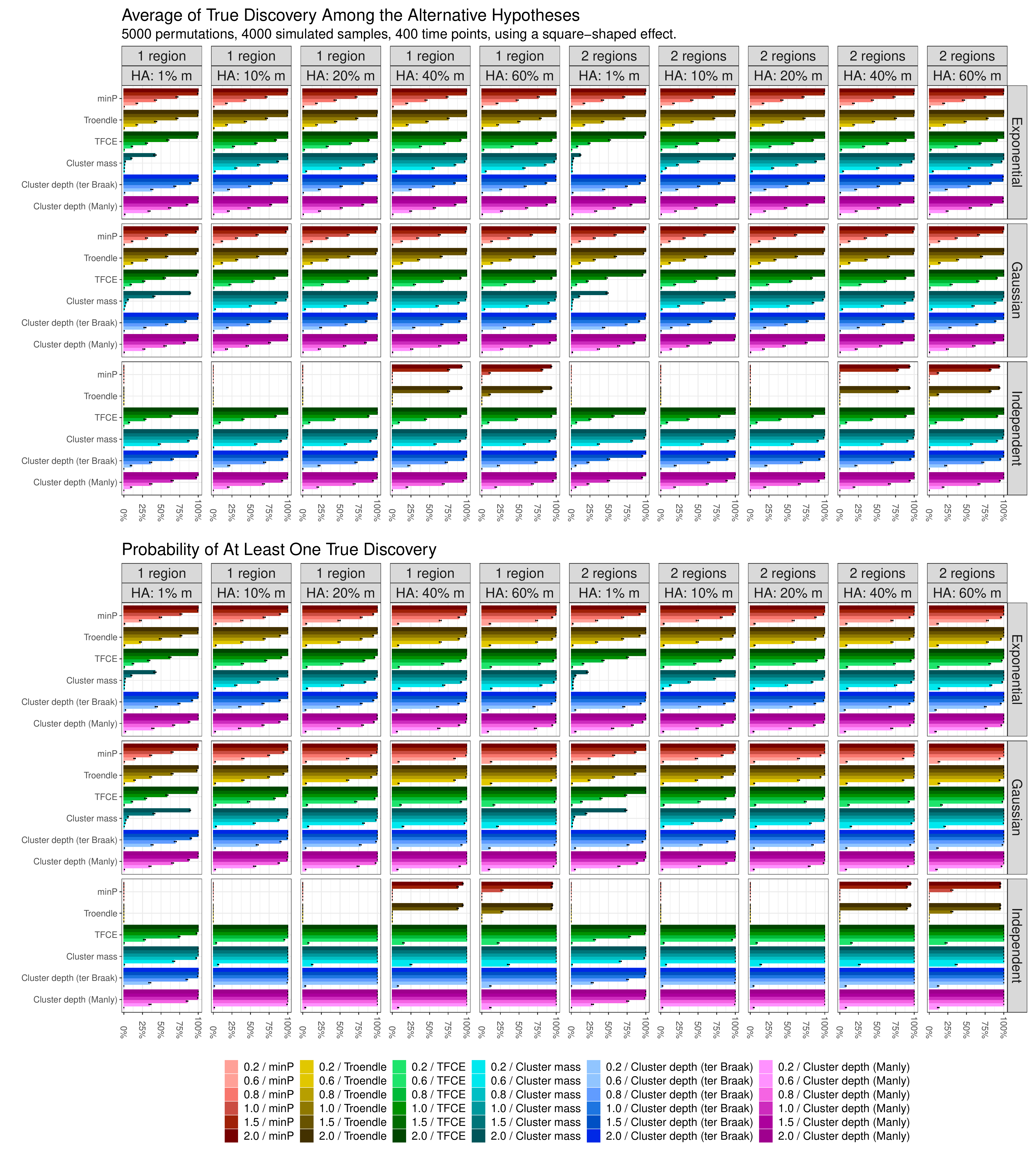}
\caption{Average and disjunctive power for the same simulation settings presented in Figure~\ref{fig:fwer_sq_ok}. Different numbers and lengths of true effect regions are presented in columns, different noise structures in rows, the different multiple comparisons procedures in colors, and the different effect sizes of the true effects in color luminance.  In all these simulation settings, the cluster depth tests have an outstanding power for a procedure that controls the FWER.}\label{fig:power_sq_ok}
\end{figure}

The two aspects of the empirical power in the same settings as in the two previous figures are displayed in Figures~\ref{fig:power_sq_ok} and~\ref{fig:power_tri_ok}. The cluster depth tests achieve a very high power, even larger than the cluster mass test, when the region of true effect is small. The power of the cluster depth tests is also systematically higher than the TFCE power, especially when the true effects have a small effect size/signal-to-noise ratio. This is an outstanding performance since the cluster mass and TFCE do not control the FWER and therefore should not even be displayed. Using our simulation setting, and especially when the region of true effect is small, the min-\textit{p} or \cite{troendle_stepwise_1995} procedures exhibit a much lower power than the cluster depth tests.

\begin{figure}[H]
\includegraphics[width=\textwidth, height=\textheight, keepaspectratio]{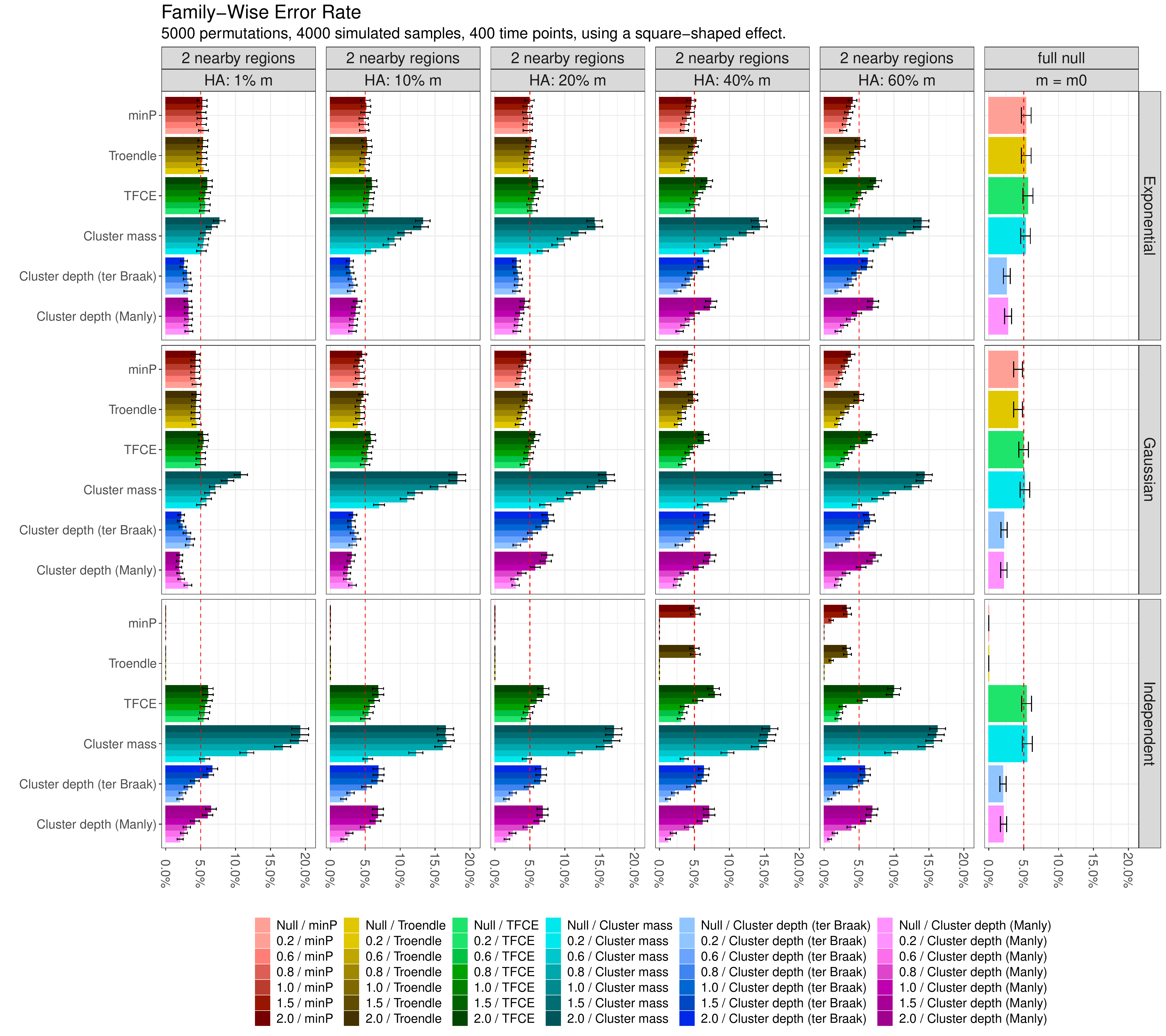}
\caption{Empirical FWER under the least favorable case for the cluster depth tests (but physiologically implausible). Two regions of true effect are present and are separated only by one single time point under the null. Different lengths of the true effect regions are presented in columns, different noise structures in rows, the different multiple comparisons procedures in colors, and the different effect sizes of the true effects in color luminance. As expected, in some cases the empirical FWER of the cluster depth tests grows above 5\%.} \label{fig:fwer_sq_f}
\end{figure}

In order to understand the limits of the proposed method, we simulated data under its least favourable case. Figure~\ref{fig:fwer_sq_f} shows the empirical FWER when the two regions of true effects are separated only by one single time point under the null. As expected, in some cases the cluster depth tests fail to keep the FWER at the 5\% level and therefore do not guarantee the strong control of the FWER when several true regions are too close. Note, however, that it is never higher than 10\% in this least favourable case.


\section{Example of Data Analysis}\label{sec:data}

\cite{tipura_attention_2019} analyzed EEG recordings from an experiment in attention shifting. The EEG signals were recorded at 1024 Hz for 614 time points (600 ms) after the presentation of an image. The 15 participants were recorded in several experimental conditions, including the \textit{visibility} of the image, either 16 ms (subliminal) or 166 ms (supraliminal); the \textit{emotion} induced by the image, either angry or neutral; and the \textit{direction} of the image on the screen, either on its left or on its right. The experimental design corresponds to a repeated measures ANOVA when averaging over the images. The permutation method from \cite{kherad-pajouh_general_2015} provides tests for any main or interaction effect in repeated measures ANOVA and can be used in combination with the multiple comparisons procedures presented in Section~\ref{sec:intro}. Similarly to the ``Manly'' permutation method, it does not asymptotically converge to a stationary distribution in presence of true effects and thus Theorem~\ref{thm:fwer} does not formally apply in this case. However, the above simulation study shows that the cluster depth tests seem robust to the permutation method, which justifies the use of the \cite{kherad-pajouh_general_2015} permutation method for repeated measures designs.

\subsection{Single Channel Analysis}
\begin{figure}[H]
\includegraphics[width=\textwidth, height=\textheight, keepaspectratio]{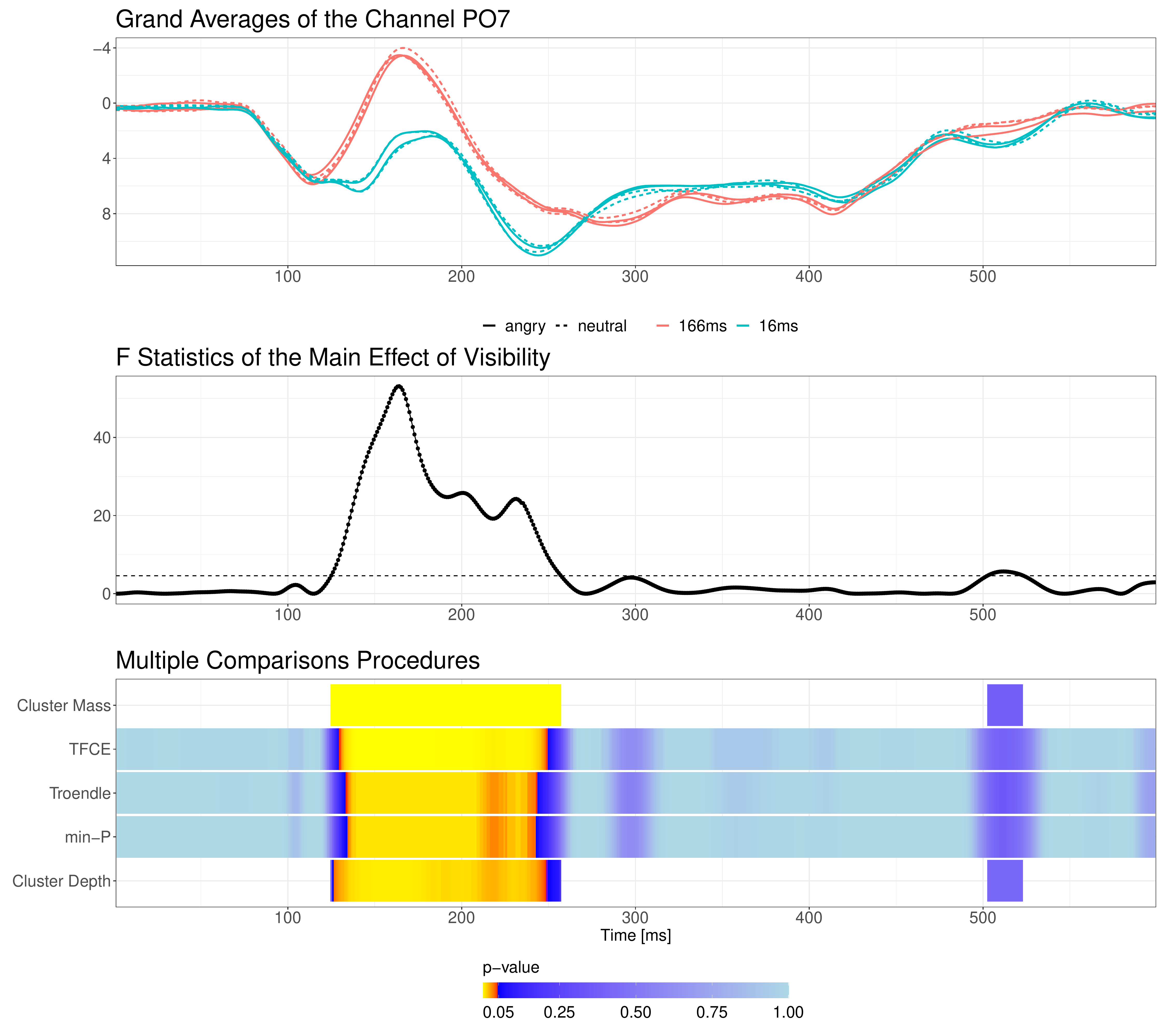}
\caption{Top Panel: Grand averages of the eight conditions of the PO7 channel from the experiment in \cite{tipura_attention_2019} ( 614 time points or 600 ms). Middle Panel: F test of the main effect of visibility. Bottom Panel: Multiple comparisons procedures. All methods detect one cluster. However, the cluster depth tests detect the largest number of time points while taking into account the false positives at the border of the cluster.}\label{fig:real_data}
\end{figure}

Figure~\ref{fig:real_data} shows the results of several multiple comparisons procedures for the main effect of the visibility of the P07 channel using 6000 permutations.

Using the cluster mass test produces a single large significant cluster that begins at the 128th measure (125ms) and ends at the 263th measure (257ms) or as soon as the statistic crosses the threshold. However, the cluster mass test does not give evidence that the first and last time points of the significant cluster are different from noise. The significant cluster may be the combination of a cluster created by noise and a true effect as in Figure~\ref{fig:cm_problem}, Panel ``Head'', ``Tail'' or ``Both''.

When using the cluster depth tests, we found that the first two cluster depths of the observed cluster are typical of one created from noise only. Therefore, we do not reject the null hypothesis for these two time points. However, from time point 130 (127ms), corresponding to the third cluster depth, we observed a statistic that is uncommon for a cluster created from noise only, and we reject its null hypothesis. At the end of the cluster, we cannot reject the null hypotheses for the ten last time points, from 254 to 263 (248ms to 257ms). It means that these ten time points are typical of the tail of clusters created from noise only. The cluster depth tests allow us to claim, with a known error rate of $\alpha_{\rm FWER}$, that the true difference between experimental conditions starts no later than 127ms and ends no earlier than 248ms.
Troendle's procedure can be interpreted point-wise as well. It finds a significant difference from time point 137 (134ms) to 249 (243ms). Troendle's procedure is always more powerful that the min-\textit{p} procedure and, for this dataset, the former discovers two additional time points compared to the latter (not displayed). Finally, the TFCE gives a significant $p$-value within time points 133 (130ms) to 255 (249ms). However, we should not have a point-wise interpretation of the TFCE, despite its relatively good results in our simulation study.

\subsection{Multi-Channels Analysis}\label{sec:datamultichannel}

On the same dataset, we consider the multiple comparisons problem over all 64 channels and all 614 time points. The cluster mass test and the TFCE can be generalized to higher dimension whenever a sensible adjacency between tests can be defined \citep{maris_nonparametric_2007}. Thereby, we need to define an adjacency matrix of the channels, on the basis of their spatial adjacency. For instance, two channels are defined as adjacent when their Euclidean distance does not exceed some predefined value $\delta$. The resulting adjacency matrix must form a connected graph. Using channel- and time-adjacency, we can define clusters in space and time and perform the cluster mass test or TFCE. As discussed in Section~\ref{sec:multi}, the cluster depths on multiple channels do not directly use adjacency, although the correlation of nearby statistics will influence the distributions and the procedure. The multi-channels extension of the cluster depth tests is implemented in the \textit{permuco4brain} R package \citep{frossard_permuco4brain_2021}.

\begin{figure}[H]
\includegraphics[width=\textwidth, height=\textheight, keepaspectratio]{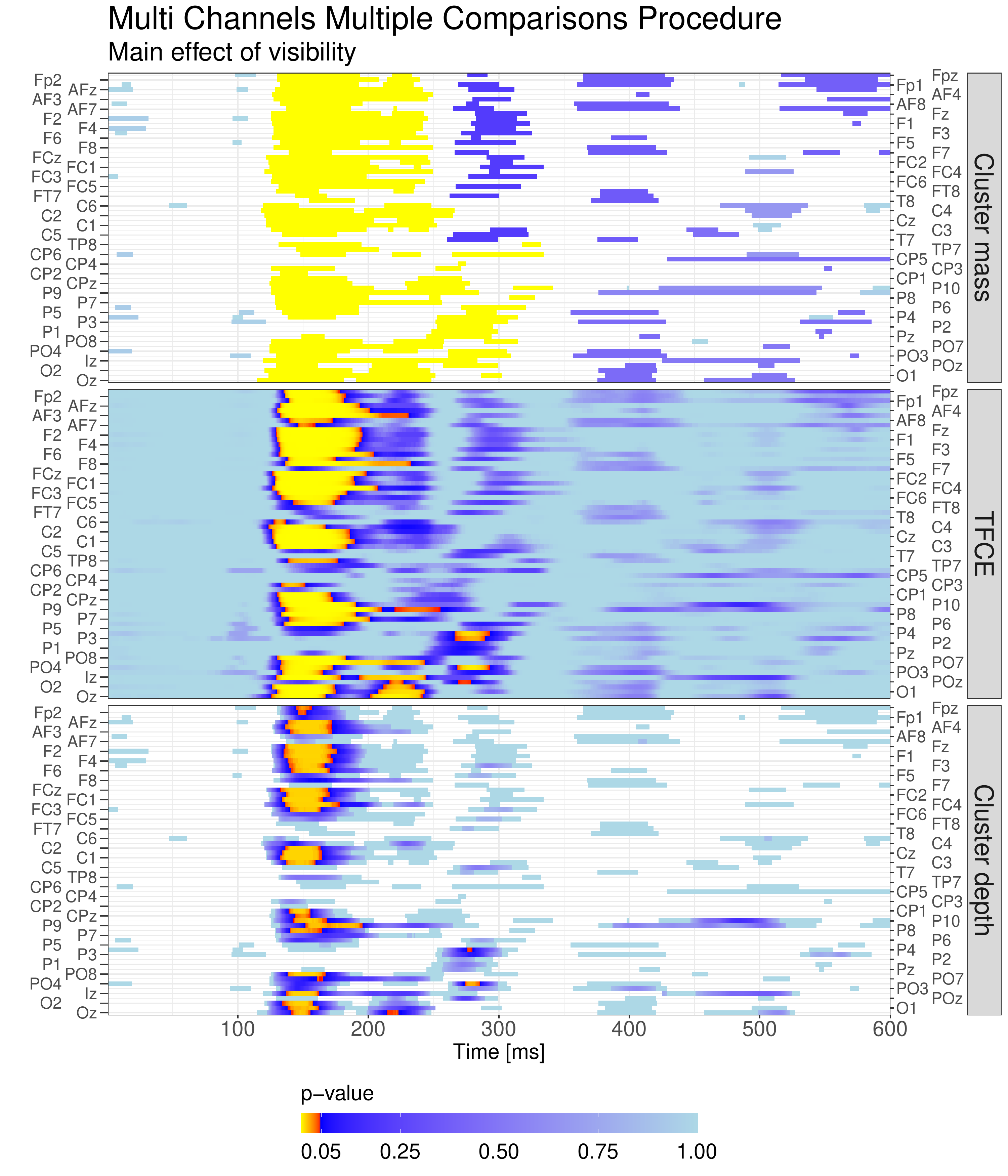}
\caption{Multi-channels analysis of the data presented by \cite{tipura_attention_2019} using 6000 permutations. The channels are presented in rows and the time in columns. The color scale represents the $p$-values. The cluster mass test, the TFCE and the cluster depth tests discover some differences in visibility. However, only the cluster depth tests allow a point-wise and channel-wise interpretation of the significant results.}\label{fig:multichannel}
\end{figure}

As the adjacency spans in both space and time, the number of neighbors of a region of true effects is not capped at two and may be much larger. For the TFCE and cluster mass test each additional neighbor of a region of true effect increases the probability of a false discovery. When interpreting the results of the cluster mass test and TFCE in Figure~\ref{fig:multichannel}, we cannot conclude, with a 5\% error rate, that all the significant time points are true effects; we cannot even conclude on the location of the border of the significant region or on a given channel. The error rate is certainly much higher for the cluster mass test than for the TFCE, but there is nevertheless no guarantee for the TFCE. Using the cluster depth tests, researchers can interpret the starting and ending time points of all discovered clusters for each channel while controlling the (global) error rate. 


Let us note that when controlling for all 64 channels, the multi-channels cluster depth tests provide a more stringent correction than the single channel analysis, since we discover a true difference on the channel PO7 that begins no later than time point 165 (161ms) and ends no earlier than time point 169 (165ms).

In this example, we performed $64\times 614 = 39296$ tests and it creates a numerical issue when using the min-\textit{p} or Troendle method with ``only'' 6000 permutations. This problem already appeared in the simulation of Section~\ref{sec:sim}, on independent tests. The number of permutations needed to obtain valid results for the min-\textit{p} and Troendle is probably much bigger that 6000 and the computation burden for this dataset is probably above realistic resources.

\section{Conclusion}

The cluster depth tests are a new multiple comparisons procedure for the massively univariate tests. We show that they are closely related to the cluster mass test and to the procedure of \cite{troendle_stepwise_1995}. Using simulation, we show that it is both powerful and controls the FWER even in the presence of physiologically plausible true effects. Moreover, we show that the cluster depth tests have two main advantages in comparison to the TFCE or the cluster mass test: they allow a point-wise  and channel-wise interpretation and can be used for the timing of effects in M/EEG.

\section{Acknowledgment}
Visualizations are made using the R packages, ggplot2 \citep{wickham_ggplot2_2016}, patchwork \citep{pedersen_patchwork_2020} and ggrepel \citep{slowikowski_ggrepel_2020}.

We thank Eda Tipura for sharing the EEG data presented in Section~\ref{sec:data}.

The simulations were performed using the ``Baobab'' HPC at the University of Geneva.

\printbibliography

\clearpage  
  
\appendix

\counterwithin{figure}{section}

\section{Proof of Theorem~\ref{thm:fwer} and aspects for other methods}
\label{sec:theoryappendix}

\subsection{Proof of Theorem~\ref{thm:fwer}}
\label{sec:proof}

Since removing the group averages (or more generally using the \cite{terbraak_permutation_1992} method) implies working on the residuals (from the full model), if the errors $\epsilon_s$ are stationary, the residuals will be asymptotically stationary. Furthermore, when using a $t$ or a $F$ statistic, as asserted by \cite{terbraak_permutation_1992}, ``the test statistic is asymptotically pivotal \cite{hall_effect_1989}: for normal errors, the distribution of the $F$-ratio does not depend on [the parameters] and for nonnormal errors this will be asymptotically true.'' This ensures that the joint distribution of the permuted statistics, the statistical signal, converges asymptotically to the true joint distribution under the full null hypothesis.

For each cluster depth $j$, the procedure and justification for constructing the distribution of $D_{\textrm{depth: }j}$ is similar to the one for the cluster mass test: the permuted signals provide a genuine way to construct the null distribution for the chosen statistic. For the depth from the head, what plays the role of the cluster statistic is the value of the statistic from the $j^{\mbox{th}}$ time point from the head (in clusters whose length is at least $j$ and that do not start at the first time point). Regardless of the number of true null and alternative (i.e., regardless of the values of the $\beta_s$), the chosen permutation scheme provides an asymptotically genuine distribution of cluster depth $j$ as if the signal was under the full null.

Using this procedure for all possible cluster depths from 1 to $m-1$ not only provides null distributions for these $m-1$ tests but also their joint distribution under the null. Any legitimate multiple comparison correction would then correct for these $m-1$ tests, and would work, but \cite{troendle_stepwise_1995} is probably the one that most wisely uses the constructed joint distribution and is probably the most powerful.


For any time point $s$ that belongs to a cluster, let $p_{sH}$ be the $p$-value from the head, as obtained at the end of Algorithm~\ref{algo:all}, i.e., after \cite{troendle_stepwise_1995} correction for multiple tests.
We will call it the corrected $p$-value from the head. For the tail, we similarly define $p_{sT}$. The final $p$-value of Algorithm~\ref{algo:both} is simply $p_{sB} = \max \{p_{sH}, p_{sT}\}$.

Under the assumption that all clusters contain at most one bloc of true effects, we can split each cluster in three oracular intervals $A$, $B$ and $C$ such that all time points in $A$ and $C$ are under the null hypothesis and all time points in $B$ are under the alternative hypothesis (i.e., the bloc of true effects). If the effect is at a border, a given cluster may contain no $A$ or $C$ region and if a cluster contains no true effect, all time points will be in interval $A$. Now, for all time points within an $A$ interval, we define the oracular $p$-value $p_{sO} := p_{sH}$, and for all time points within a $C$ interval, we define $p_{sO} := p_{sT}$. 
Let's now show that the FWER is controlled at level $\alpha_{\rm FWER}$:

\begin{eqnarray}
  \nonumber
  \mbox{FWER} & = & \Prob \{ \mbox{at least one Type I error in one of the $A$ or $C$ intervals} \}
  \\
  & = &  \Prob \{ \mbox{smallest $p_{sB}$ is smaller than $\alpha_{\rm FWER}$ over all $A$ and $C$ intervals} \}
  \nonumber
  \\
    \label{eq:pf_fwer}
  & \leq &  \Prob \{ \mbox{smallest $p_{sO}$  is smaller than $\alpha_{\rm FWER}$ over all $A$ and $C$ intervals} \},
\end{eqnarray}
where the last inequality stems from the fact that the oracular $p$-value $p_{sO}$ is smaller than the final $p$-value $p_{sB}$ of Algorithm~\ref{algo:both}. Note that similarly to many proofs concerning the FWER, controlling for the most extreme value (min or max) will lead to the control over all tests. Let $p_{min}$ be the smallest $p_{sO}$ (over all $A$ and $C$ intervals) and let $s_{min}$ be its corresponding time point. Suppose that $s_{min}$ is in an $A$ interval, and let $j_{min}$ be its position in the cluster, i.e., its cluster depth from the head. By construction, $s_{min}$ is under the null hypothesis, and – critically – all prior time points in the cluster are also under the null hypothesis. Its $p$-value $p_{s_{min}O}$, which is equal to $p_{s_{min}H}$, is constructed from the cluster depth $j_{min}$ from the head. The chances that a time point in this situation (under the null and all leading points within the cluster also under the null) will obtain a $p$-value $p_{sH}$ smaller than $\alpha_{\rm FWER}$ is bounded by $\alpha_{\rm FWER}$ since the joint distribution of cluster depths from the head is constructed on $m$ tests and is corrected for multiple comparisons. This shows that in that case, the probability in Equation~\ref{eq:pf_fwer} is bounded by $\alpha_{\rm FWER}$ (note that it is even conservative, as the number of time points under the null is larger in the permuted data than in the original data).

By contrast, if $s_{min}$ is in a $C$ interval, its oracular $p$-value is $p_{sT}$, i.e., stemming from the cluster depth from the tail. A symmetric reasoning can be done on the trailing points, which also lead to the above bound. Since by assumption null tests can only be in $A$ or $C$ intervals, this proves the theorem.

\section{Additional Simulation Results}

 \begin{figure}[H]
\includegraphics[width=\textwidth, height=\textheight, keepaspectratio]{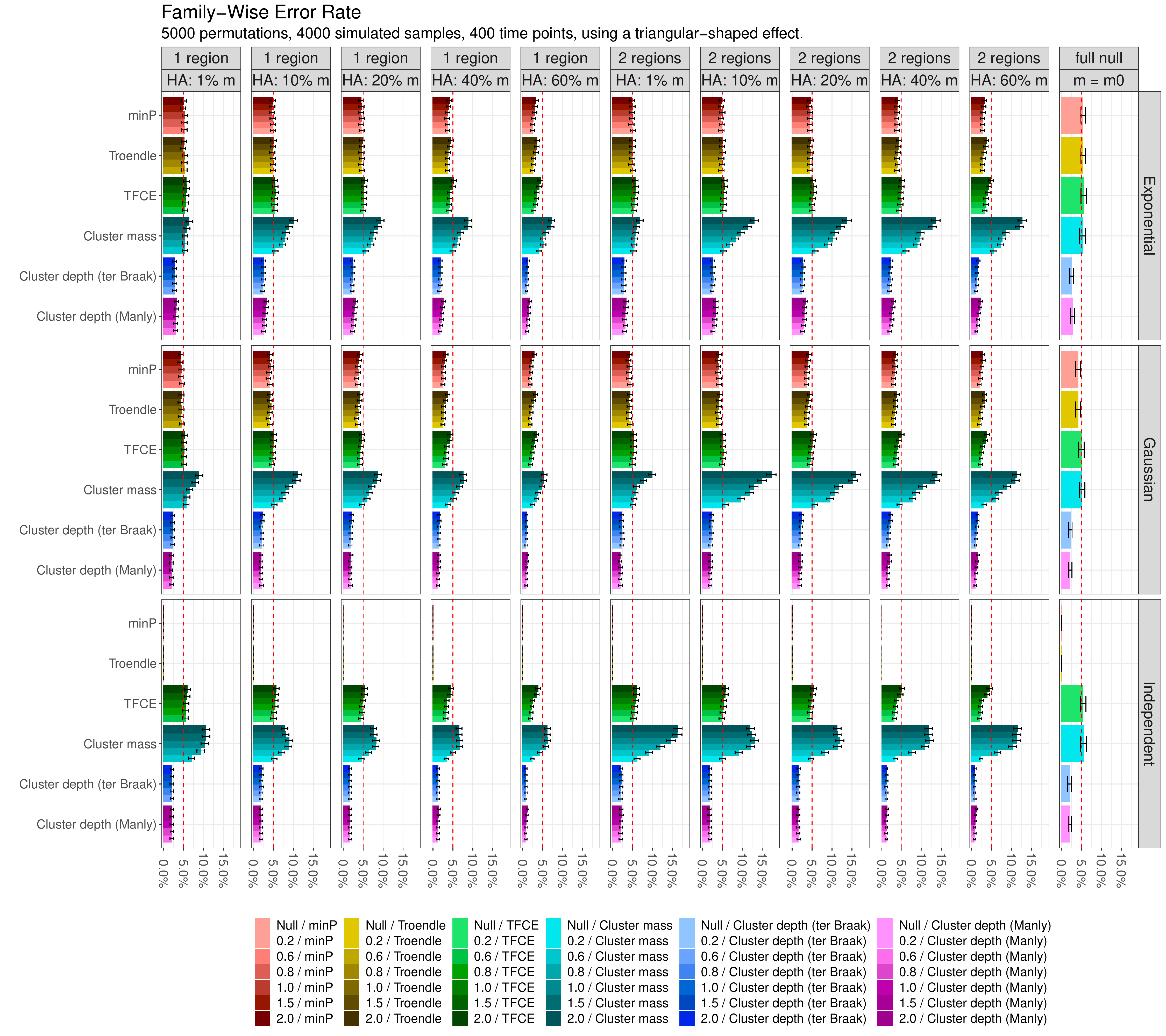}
\caption{Empirical FWER in the same settings as in Figure~\ref{fig:fwer_sq_ok}, except with a triangular-shaped effect, to evaluate the influence of a continuously growing effect on the FWER. In these simulation settings, the cluster depth tests control the FWER, in a conservative way.} \label{fig:fwer_tri_ok}
\end{figure}

 \begin{figure}[H]
\includegraphics[width=\textwidth, height=\textheight, keepaspectratio]{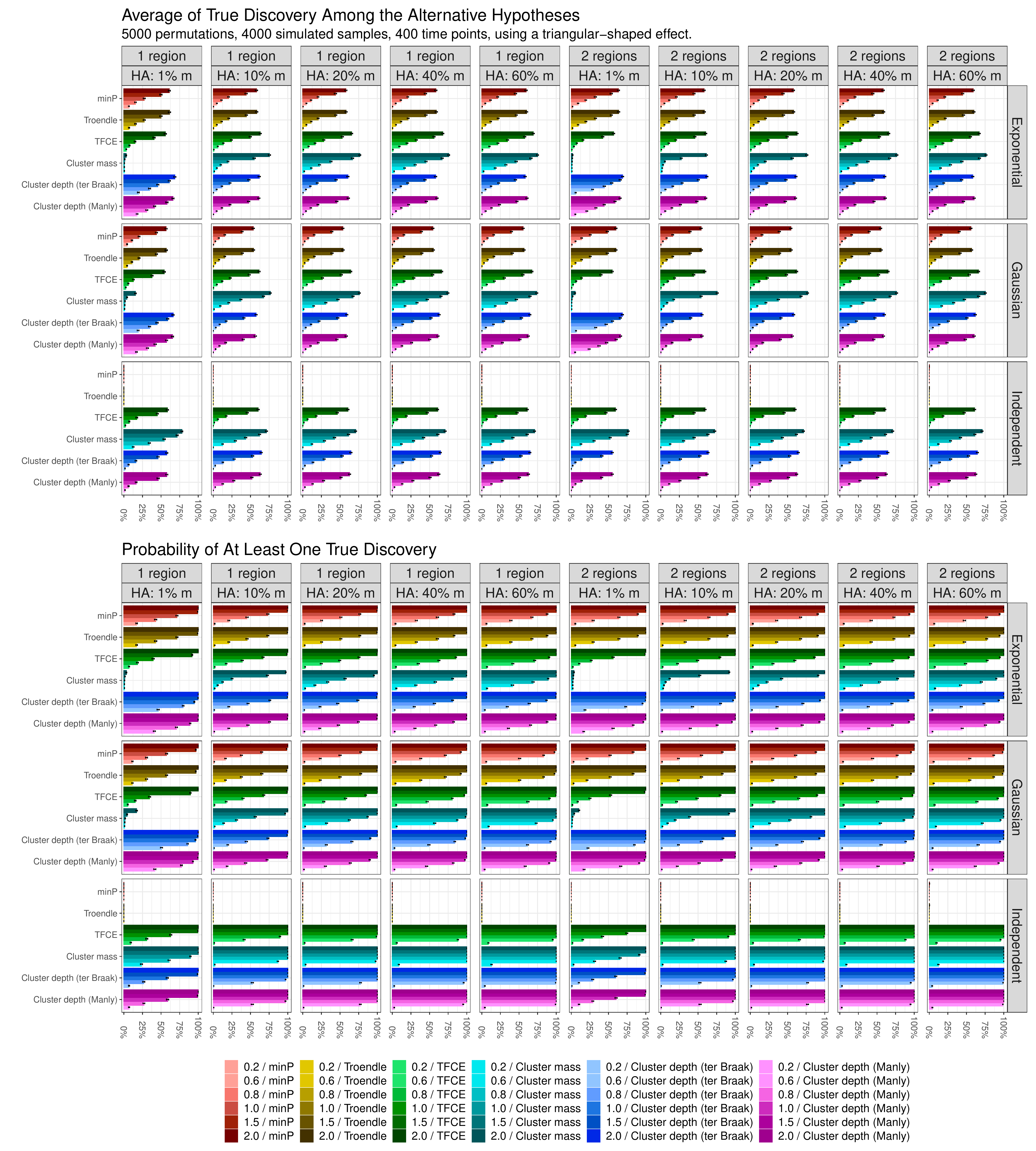}
\caption{Average and disjunctive power for the similarly to \ref{fig:power_sq_ok}, except for a triangular-shaped effect. The cluster depth tests achieve a large power while controlling the FWER.}\label{fig:power_tri_ok}
\end{figure}

\end{document}